\documentstyle[12pt]{article}
\input axodraw.sty
\begin{document}
\title{Infrared behavior of graviton-graviton scattering}
\author{\normalsize John F. Donoghue$^1~$ {\small\it and}~~Tibor Torma$^2$\\[1ex]
  \small\em$^1$Department of Physics and Astronomy, University of Massachusetts\\
  \small\em Amherst, MA 01003, USA\thanks{donoghue@phast.umass.edu}\\[.5ex]
  \small\em$^2$Department of Physics, University of Toronto\\
  \small\em Toronto, Ontario, Canada M5S1A7\thanks{kakukk@physics.utoronto.ca}}
\maketitle
\begin{picture}(0,0)
\put(340,270){Preprint UTPT-99-01}
\end{picture}
\begin{abstract}
The quantum effective theory of general relativity,
independent of the eventual
full theory at high energy, expresses graviton-graviton
scattering at one loop order ${\cal O}\left(E^4\right)$ with
only one parameter, Newton's constant. Dunbar and Norridge
have calculated the one loop amplitude using string based
techniques.  We complete the calculation by showing that the
$1\over d-4$ divergence which remains in their result comes
from the infrared sector and that the cross section is finite and
model independent when the usual bremsstrahlung diagrams are
included.
\end{abstract}
\baselineskip 19pt
\section{Introduction}

The simplest low energy process in quantum gravity
is graviton-graviton scattering. Although experimentally
unobservable, this reaction forms an interesting theoretical
laboratory that illustrates the workings of quantum gravity. If
general relativity is the correct low energy classical theory of
gravity, then its quantum theory forms an effective field theory
capable of analyzing the low energy quantum effects.
Graviton-graviton scattering is particularly useful in
illustrating the logic of predictions in a quantum effective
theory. Indeed, at one loop order this reaction provides a
model-independent {\it quantum} prediction of general
relativity.\label{predict}

At tree level, the graviton-graviton scattering amplitude is
simple in the helicity basis, although the calculation to obtain
this result from the Einstein action is not so simple.  With
$+\,(-)$ representing helicity~$+2\,(-2)$\footnote{Note that in
our notation crossing also requires one to flip the $\pm$ sign for
the affected gravitons. This implies in particular that
${\cal A}(- -;++)$ must be a symmetric function of $s,t$
and $u$.}, all tree amplitudes for $1+2 \rightarrow 3+4$
vanish except those related to ${\cal A}^{tree}(++;++)$
by crossing and~\cite{tree}\label{refadded:tree}
\begin{equation}\label{eq:1}
{\cal A}^{tree}(++;++)  =  {i\over4}\,{\kappa^2 s^3 \over t u}
\end{equation}
Here  $\kappa^2 =32\pi G$ and $s=(p_1 + p_2)^2,
\ t=(p_1-p_3)^2,\ u=(p_1-p_4)^2$
denote the usual Madelstam variables.

It is  simple to  show that graviton-graviton scattering should
be finite and parameter independent at one-loop order~\cite{tHV}.
In the
effective low energy theory~\cite{graveff} gravitational effects
are expanded in a derivative expansion with all terms satisfying
general covariance
\begin{equation}\label{eq:lagr}
S_{grav}=\int d^4x\, \sqrt{g}\, \left[\,{2\over{\kappa^2}}\, R+c_1\, R^2 +
c_2\, R_{\mu \nu} R^{\mu \nu} + \ldots + {\cal L}_{matter}\,\right]
\end{equation}
\noindent Here $\kappa^2=32 \pi G$ and $G$ is
Newton's constant, $c_{1,2}$ are unknown dimensionless
parameters which contain information about the (presently
unknown) ultimate high energy theory.  A third covariant of
order~$R^2$, $R_{\mu \nu \alpha \beta} R^{\mu \nu \alpha
\beta}$ can be removed in four dimensional space-time
through the use of the Bianchi identities.  Since the curvature
involves two derivatives of the metric, the Einstein action (the
term with $R$) is seen to be of order $E^2$, while
$R^2$ and $R_{\mu \nu} R^{\mu \nu}$ are of order $E^4$.

Loop diagrams obey a power-counting theorem~\cite{thv,dt}.
One loop diagrams formed from vertices given by the Einstein
action yield effects at order~$E^4$ -- any process with more
loops is higher order in the energy expansion. The ultraviolet
divergences at one loop necessarily have the same structure as
the local Lagrangian in Eqn.~(\ref{eq:lagr}), which means that
they must be proportional to $R^2$ or
$R_{\mu\nu}R^{\mu\nu}$. Then, at this order, the ultraviolet
divergences can be absorbed into renormalized values of the
the parameters $c_{1,2}$. These renormalized constants are
unknown and will be different depending on the nature of the
theory that forms the ultimate correct high energy theory which
includes gravity. In this sense these parameters are model
dependent. However, they do not contribute to the process of
graviton-graviton scattering. At the order that we are working,
the $R^2$ Lagrangians are applied to form vertices for on-shell
amplitudes, which is to say that the equations of motion are
satisfied for the external states. However, the equations of
motion for the purely gravitational sector are
$R_{\mu\nu}= 0$, and hence $R=0$ also. Thus the effects of
both of the $R^2$ terms in~Eqn.~(\ref{eq:lagr}) vanish in
purely gravitational processes. It is this argument that tells us
that graviton-graviton scattering is finite and independent of
any unknown parameters at one loop order.

The power counting theorem is manifest in the one-loop results
calculated by Dunbar and Norridge~\cite{DunNor}.
The one-loop amplitude is formed by using the lowest order
tree amplitude twice in order to produce a loop diagram, and
hence carries coupling constants $\kappa^4 \sim G_N^2$.
Dimensionally this requires that the result carry four powers of
the external energies. This is seen in the results:
\begin{eqnarray}\label{eq:2}
{\cal A}^{1-loop}(++;--) & = & -i\,{\kappa^4 \over 30720
\pi^2}
\left( s^2+t^2 + u^2 \right)   \nonumber\\
{\cal A}^{1-loop}(++;+-) & = & -{1 \over 3}
{\cal A}^{1-loop}(++;--)\nonumber \\
{\cal A}^{1-loop}(++;++) & = &\frac{\kappa^2}
{4(4\pi)^{2-\epsilon}}\,
 \frac{\Gamma^2(1-\epsilon)\Gamma(1+\epsilon)}
 {\Gamma(1-2\epsilon)}\,
 {\cal A}^{tree}(++;++)\,\times(s\,t\,u)\\
&&\hspace{-0em}\times\left[\rule{0pt}{4.5ex}\right.
\frac{2}{\epsilon}\left(
\frac{\ln(-u)}{st}\,+\,\frac{\ln(-t)}{su}\,+\,\frac{\ln(-s)}{tu}
\right)+\,\frac{1}{s^2}\,f\left(\frac{-t}{s},\frac{-u}{s}\right)
\nonumber\\&&\hspace{1.4em}
+2\,\left(\frac{\ln(-u)\ln(-s)}{su}\,+\,\frac{\ln(-t)\ln(-s)}{tu}\,+\,
\frac{\ln(-t)\ln(-s)}{ts}\right)
\left.\rule{0pt}{4.5ex}\right]\nonumber
\end{eqnarray}
where
\begin{eqnarray}\label{eq:f}
f\left(\frac{-t}{s},\frac{-u}{s}\right)&=&
\frac{(t+2u)(2t+u)\left(2t^4+2t^3u-t^2u^2+2tu^3+2u^4\right)}
{s^6}
\left(\ln^2\frac{t}{u}+\pi^2\right)\nonumber\\&&
+\frac{(t-u)\left(341t^4+1609t^3u+2566t^2u^2+1609tu^3+
341u^4\right)}
{30s^5}\ln\frac{t}{u}\nonumber\\&&
+\frac{1922t^4+9143t^3u+14622t^2u^2+9143tu^3+1922u^4}
{180s^4},
\end{eqnarray}
\noindent and all logarithms with negative arguments are
understood to have a $-i\pi$ imaginary part. Note that this
represents a real {\it tour-de-force}. Done in conventional field
theory, the calculation is formidably difficult. It is a tribute
to the string based techniques that the results are obtainable
with less then Herculean effort. Indeed, after calculating the
graviton loops, the authors write down the result for massless
scalars, fermions and photons in the loops in just a few lines.
However, the result is not tied to the validity of string theory as
a fundamental theory -- the technique is simply an efficient
way to calculate the results of usual (quantum) general
relativity.

One notices that the one loop amplitude in Eqn.~(\ref{eq:2})
contains a factor of $\frac{1}{\epsilon}$, i.e. it is {\it not}
finite. At first, this seems to contradict the general reasoning
given above. However, in the complete calculation of the
physical process of graviton scattering, there will also be
bremsstrahlung diagrams describing the radiation of soft
gravitons off the external graviton lines. When calculated in a
$d=4-2\epsilon$ dimensional phase space these infrared effects
also bring in a
$\frac{1}{\epsilon}$ factor. If the divergence
in~Eqn.~(\ref{eq:2}) is an infrared divergence, and if the
effective field theory of gravity behaves as a proper effective
field theory, then the infrared loop effects should be canceled
against the soft radiation. While there are good reasons for
believing that the gravitational effective field theory should be
well behaved in the infrared, the long-standing doubts about
quantum gravity make it worthwhile to check this property in
the only complete calculation available. In~Ref.~\cite{weinb}
it was shown that the scattering of spin-0 fields is infrared
finite even in the limit when their masses vanish. However, it
was only conjectured there that the same is true for massless
matter of higher spin (a situation similar to graviton-graviton
scattering). One also notes that the scale of the logarithm is not
defined. This is an indication that the calculation is incomplete.
We will see that the scale in the logarithm comes from an
infrared regulator for soft gravitons. Finally, part of our
motivation comes from a minor quibble with the argument
given above. In the effective Lagrangian we removed
the~$(R_{\mu\nu\alpha\beta})^2$ term by the use of an
identity that is only valid in exactly four dimensions. Indeed, in
any higher dimension the argument given would not apply, and
the graviton scattering amplitude would contain a model
dependent parameter. This means that in
the quantum theory we can only be certain of the result if we
use a regularization scheme that works in four dimensions.
However, the only scheme that we know about that preserves
the symmetries of general relativity is dimensional
regularization, and it was that scheme used in
Ref.~\cite{DunNor}. While it is unlikely that the regularization
scheme would lead to an extra divergence, we also want to
confirm that the residual divergence is not an artifact of the
ultraviolet regularization.

\section{Soft gravitons in graviton-graviton scattering}

We will explicitly calculate the divergences in the one-loop
differential cross section for graviton-graviton scattering.
We will find a complete cancellation of infrared divergences
when we calculate the cross section up to ${\cal
O}(\kappa^6)$, including the Bremsstrahlung graphs, as
shown in Fig.~\ref{fig:gggen}.

\begin{figure}[thb]
\begin{center}
\begin{picture}(320,155)
 \put(0,85){
  \begin{picture}(90,70)
   \put(3,24){$\kappa^2\times$}
   \Line(24,35)(38,35)
   \Line(24,20)(38,20)
   \Line(68,35)(82,35)
   \Line(68,20)(82,20)
   \CArc(53,27.5)(16.77,0,360)\put(45,19){\Huge {\cal A}}
   \put(91,21){\huge+}
   \put(126,0){
    \put(-13,24){$\kappa^4\times$}
    \begin{picture}(90,70)
    \Line(8,35)(18,35)
    \Line(8,20)(18,20)
    \Line(48,35)(82,35)
    \Vertex(60,20)3\Line(48,20)(82,20)\Vertex(60,35)3
    \Photon(60,20)(60,35)2 5
    \CArc(33,27.5)(16.77,0,360)\put(25,19){\Huge {\cal A}}
    \end{picture}}
   \put(217,21){\huge+(etc.)}
   \Line(1,0)(1,55)\Line(291,0)(291,55)\put(295,53){\huge2}
   \put(306,21){\huge+}
  \end{picture}}
 \put(90,0){
   \put(-46,21){\huge+}
   \begin{picture}(90,70)
    \Line(8,35)(18,35)
    \Line(8,20)(18,20)
    \Line(48,35)(82,35)
    \Vertex(60,20)3\Line(48,20)(82,20)
    \Photon(60,20)(82,27.5)2 5
    \CArc(33,27.5)(16.77,0,360)\put(25,19){\Huge {\cal A}}
    \put(91,21){\huge+(etc.)}
    \put(-13,24){$\kappa^3\times$}
    \Line(-15,0)(-15,55)\Line(165,0)(165,55)\put(168,53){\huge2}
   \end{picture}}
\end{picture}
\end{center}
\caption{\label{fig:gggen}The expansion of the cross section
in {$\kappa$} in graviton-graviton scattering. The quantity
${\cal A}^{tree}$ represents the sum of all tree level diagrams.
Solid lines represent hard gravitons, wavy  lines are soft
gravitons.}
\end{figure}
In this figure we explicitly show all factors of $\kappa$.
The first term in the figure (and five additional graphs, not
shown, with graviton exchanges between various pairs of
external legs) is already included in the full one-loop scattering
calculation and has an infrared divergence. This divergence is
canceled by another divergence in the second term in the
figure, a soft Bremsstrahlung process, which should be added
as it is degenerate in energy with pure hard scattering. The
second line in the figure shows that the actual cancellation
occurs in the ${\cal O}(\kappa^6)$ terms because the leading
${\cal O}(\kappa^4)$ is tree level and infrared finite.

We will derive a general formula for the infrared divergences which uses the on-shell Born
amplitude. The most convenient regularization procedure is
dimensional regularization. We calculate the IR
divergent part of the graviton radiation
term in Fig.~\ref{fig:gggen} and show
[cf. Eqn.~(\ref{eq:cross-ggfin})] that
to do so we only need to know the on-shell tree level
amplitudes${\cal
A}^{tree}(\lambda_1,\lambda_2,\lambda_3,\lambda_4)$. We
always work in the helicity basis and $\lambda_i=\pm$ stands
for the helicity of the $i^{th}$ hard graviton. The only
divergence occurs when the gravitons have helicity
assignments $++;++$ (and in the cases related to this by
crossing) and that is the only case when the tree level
amplitude is nonvanishing. In the following we show
[Eqn.~(\ref{eq:a-part})] that soft graviton radiation does not
flip the hard graviton spins so that all IR divergences are
proportional to the tree amplitude with the same helicity.

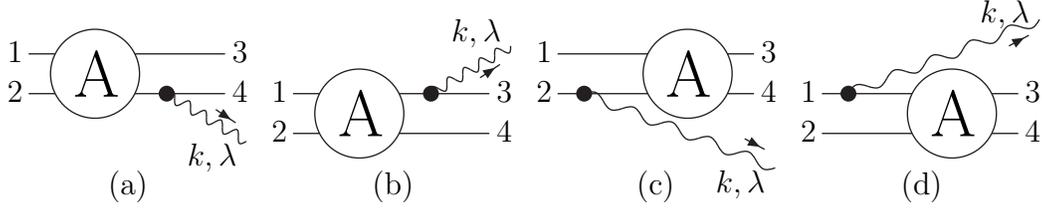
\begin{figure}[hbt]
\begin{center}
\begin{picture}(390,85)
\put(0,0){\begin{picture}(90,80)
  \put(68,23){$k,\lambda$}
  \Line(8,65)(18,65)\put(0,62){1}
  \Line(8,50)(18,50)\put(0,47){2}
  \Line(48,65)(82,65)\put(85,62){3}
  \Vertex(60,50)3\Line(48,50)(82,50)\put(85,47){4}
  \Photon(60,50)(90,32)2 5\ArrowLine(78.5,44)(86,39)
  \CArc(33,57.5)(16.77,0,360)\put(25,49){\Huge {\cal A}}
  \put(38,12){(a)}
  \end{picture}}
\put(100,0){\begin{picture}(90,80)
 \put(68,70){$k,\lambda$}
  \Line(8,50)(18,50)\put(0,47){1}
  \Line(8,35)(18,35)\put(0,32){2}
  \Line(48,50)(82,50)\put(85,47){3}
  \Vertex(60,50)3\Line(48,35)(82,35)\put(85,32){4}
  \Photon(60,50)(90,68)2 5\ArrowLine(78.5,56)(86,61)
  \CArc(33,42.5)(16.77,0,360)\put(25,34){\Huge {\cal A}}
  \put(38,12){(b)}
  \end{picture}}
\put(200,0){\begin{picture}(90,80)
  \put(68,13){$k,\lambda$}
  \Line(8,65)(42,65)\put(0,62){1}
  \Line(8,50)(42,50)\put(0,47){2}
  \Line(72,65)(82,65)\put(85,62){3}
  \Vertex(18,50)3\Line(72,50)(82,50)\put(85,47){4}
  \Photon(18,50)(90,22)2 5\ArrowLine(78.5,33)(86,29)
  \CArc(57,57.5)(16.77,0,360)\put(49,49){\Huge {\cal A}}
  \put(38,12){(c)}
  \end{picture}}
\put(300,0){\begin{picture}(90,85)
  \put(68,77){$k,\lambda$}
  \Line(8,50)(42,50)\put(0,47){1}
  \Line(8,35)(42,35)\put(0,32){2}
  \Line(72,50)(82,50)\put(85,47){3}
  \Vertex(18,50)3\Line(72,35)(82,35)\put(85,32){4}
  \Photon(18,50)(90,78)2 5\ArrowLine(78.5,68)(86,72)
  \CArc(57,42.5)(16.77,0,360)\put(49,34){\Huge {\cal A}}
  \put(38,12){(d)}
  \end{picture}}
\end{picture}
\end{center}
\caption{\label{fig:gggraphs}The four Feynman diagrams that
contribute to soft graviton radiation at lowest order in hard
graviton-graviton scattering.}
\end{figure}

The amplitude with one soft graviton radiation is the sum
of the four diagrams in~Fig.~\ref{fig:gggraphs}.
We first calculate the contribution from 2(a)
\begin{equation}\label{eq:contr-a}
{{\cal A}^{rad}_{(a)}={\cal
A}^{tree}_{\mu\nu}(1,2;3,k_4+k)\,\frac{i}
{(k_4+k)^2+i\epsilon}
 \,P_{\mu\nu,{\mu'}{\nu'}}\,i\frac{\kappa}{2}
 \tau^{{\mu'}{\nu'}}_{\alpha\beta,\lambda\rho}
 \,\overline{\epsilon^{\alpha\beta}_4}\,
 \overline{\epsilon^{\lambda\rho}}}
 \end{equation}
where writing a number in the argument of the tree amplitude
means putting those lines on shell and multiplying by the
appropriate $\epsilon^{\mu\nu}$ polarization tensor. The
matrices $I$ and $P$ denote

\begin{equation}\label{eq:Inote}{I_{\alpha\beta,\gamma\delta}=
 \frac{1}{2}\left(\eta_{\alpha\gamma}\eta_{\beta\delta}+
 \eta_{\alpha\delta}\eta_{\beta\gamma}\right)}\end{equation}
and
\begin{equation}\label{eq:Pnote,Pmat}
{P_{\alpha\beta,\gamma\delta}=
I_{\alpha\beta,\gamma\delta}-\frac{1}{2}
\eta_{\alpha\beta}\eta_{\gamma\delta}.}
\end{equation}
The gauge invariance of the tree level amplitudes implies,
for $k\to0$,
\begin{equation}\label{eq:ginv2}
{k_4^\mu\,{\cal A}^{tree}_{\mu\nu}(1,2;3,k_4+k)={\cal O}(k)}
\end{equation}
and also
\begin{equation}\label{eq:ginv3}
{\eta^{\mu\nu}\,{\cal A}_{\mu\nu}(1,2;3,k_4+k)={\cal O}(k).}
\end{equation}
These restrictions can be derived as follows. Gauge invariance implies that the
on-shell amplitude is unchanged under shifting the polarization tensor by
\begin{equation}
\epsilon^{\mu\nu}\longrightarrow\epsilon^{\mu\nu}+(k^\mu\xi^\nu+\xi^\mu
k^\nu-k\cdot\xi\ \eta^{\mu\nu}),
\end{equation}
with any four-vector $\xi^\mu$, a transformation that keeps
$k_\mu\epsilon^{\mu\nu}$ zero. In order for an amplitude
$A_{\mu\nu}\epsilon^{\mu\nu}$ to be invariant under such replacement, we need for
any on shell momentum $k$
\begin{equation}
2\xi_\mu(k_\nu A^{\mu\nu})=(k\cdot\xi)A^\mu_\mu.
\end{equation}
This must hold for any $\xi$, hence we have \mbox{Eqns.
(\ref{eq:ginv2},\ref{eq:ginv3}).}
\begin{figure}[htb]
\begin{center}
\begin{picture}(390,85)(0,0)
\put(0,65){\begin{picture}(390,20)
\put(33,10){$\mu\nu$}
\put(180,10){$\Longrightarrow$}
\Photon(50,13)(120,13)3 5
\Vertex(50,13)2
\Vertex(120,13)2
\put(125,10){$\alpha\beta$}
\put(250,10){$\frac{i}{q^2+i\epsilon}\cdot
\left[\eta^\alpha_{(\mu}\eta^\beta_{\nu)}-
\frac{1}{2}\eta_{\mu\nu}\eta^{\alpha\beta}\right]$}
\end{picture}}
\put(0,5){\begin{picture}(390,60)
\put(180,17){$\Longrightarrow$}
\Photon(50,40)(120,40)3 5\put(33,40){$\alpha\beta$}
\put(125,40){$\gamma\delta$}
\ArrowLine(58,35)(67,35)\ArrowLine(112,35)(103,35)
\ArrowLine(79,20)(79,29)
\put(87,21){$k_1$}\put(60,45){$k_2$}\put(103,45){$k_3$}
\Vertex(85,40)3
\Photon(85,10)(85,40)3 5\put(80,0){$\mu\nu$}
\put(250,20){$\frac{i}{2}\kappa\cdot
\tau^{\mu\nu}_{\alpha\beta,\gamma\delta}(k_1,k_2,k_3)$}
\end{picture}}
\end{picture}
\end{center}
\caption{\protect\label{fig:grfey}The graviton propagator and
the triple gluon vertices in harmonic gauge. For an expression
of $\tau^{\mu\nu}_{\alpha\beta,\gamma\delta}$ see
Eqn.~(\protect\ref{eq:tau3}).}
\end{figure}

The graviton propagator and the triple graviton couplings are
shown in Fig.~\ref{fig:grfey}
\begin{eqnarray}\label{eq:tau3}
&&\!\!\!\!\!\!\!\!\!\tau^{\mu\nu}_{\alpha\beta,\gamma\delta}(k_1,k_2,k_3)=\\
 &&\ \ \ P_{\alpha\beta,\gamma\delta}\left[k_2^\mu k_2^\nu+(k_2-k_1)^\mu(k_2-k_1)^\nu+
   k_1^\mu k_1^\nu-\frac{3}{2}\eta^{\mu\nu}k_1^2\right]\nonumber\\
 &&\ \ +2\,{(k_1)}_\lambda {(k_1)}_\sigma\left[
   I^{\lambda\sigma,}_{\ \ \alpha\beta}I^{\mu\nu,}_{\ \ \gamma\delta}+
   I^{\lambda\sigma,}_{\ \ \gamma\delta}I^{\mu\nu}_{\ \ \alpha\beta}-
   I^{\lambda\mu}_{\ \ \alpha\beta}I^{\sigma\nu,}_{\ \ \gamma\delta}-
   I^{\sigma\nu,}_{\ \ \alpha\beta}I^{\lambda\mu}_{\ \ \gamma\delta}\right]\nonumber\\
 &&\ \ +{(k_1)}_\lambda k_1^\mu\left(
    \eta_{\alpha\beta}I^{\lambda\nu,}_{\ \ \gamma\delta}+
    \eta^{\gamma\delta}I^{\lambda\nu,}_{\ \ \alpha\beta}\right)
   +{(k_1)}_\lambda k_1^\nu\left(
    \eta_{\alpha\beta}I^{\lambda\mu,}_{\ \ \gamma\delta}+
    \eta_{\gamma\delta}I^{\lambda\mu,}_{\ \ \alpha\beta}\right)\nonumber\\
  &&\ \ -k_1^2\left(\eta_{\alpha\beta}I^{\mu\nu,}_{\ \ \gamma\delta}+
     \eta_{\gamma\delta}I^{\mu\nu,}_{\ \ \alpha\beta}\right)
     -\eta^{\mu\nu}k_1^\lambda k_1^\sigma\left(
      \eta_{\alpha\beta}I_{\gamma\delta,\lambda\sigma}+
      \eta_{\gamma\delta}I_{\alpha\beta,\lambda\sigma}\right)\nonumber\\
  &&\ +2\,k_1^\lambda\left(
     I^{\sigma\nu,}_{\ \ \alpha\beta}I_{\gamma\delta,\lambda\sigma}(k_2-k_1)^\mu
    +I^{\sigma\mu,}_{\ \ \alpha\beta}I_{\gamma\delta,\lambda\rho}(k_2-k_1)^\nu\right.\nonumber\\
&&\left.\hspace{8ex}
    -I^{\sigma\nu,}_{\ \ \gamma\delta}I_{\alpha\beta,\lambda\sigma}k_2^\mu
    -I^{\sigma\mu,}_{\ \ \gamma\delta}I_{\alpha\beta,\lambda\sigma}k_2^\nu
     \right)\nonumber\\
  &&\ \ +k_1^2\left(
    I^{\sigma\mu,}_{\ \ \alpha\beta}I_{\gamma\delta,\sigma}^{\ \ \ \nu}+
    I_{\alpha\beta,\sigma}^{\ \ \ \nu} I^{\sigma\mu,}_{\ \ \alpha\delta}\right)
    +\eta^{\mu\nu}k_1^\lambda{(k_1)}_\sigma\left(
    I_{\alpha\beta,\lambda\rho}I^{\rho\sigma,}_{\ \ \gamma\delta}+
    I_{\gamma\delta,\lambda\rho}I^{\rho\sigma,}_{\ \ \alpha\beta}\right)\nonumber\\
 &&\ \ +\left(k_2^2+(k_2-k_1)^2\right)\left(
    I^{\sigma\mu,}_{\ \ \alpha\beta}I_{\gamma\delta,\sigma}^{\ \ \ \nu}+
    I^{\sigma\nu,}_{\ \ \alpha\beta}I_{\gamma\delta,\sigma}^{\ \ \ \mu}-
    \frac{1}{2}\eta^{\mu\nu}P_{\alpha\beta,\gamma\delta}\right)\nonumber\\
  &&\ \ -\left(k_2^2\eta_{\gamma\delta}I^{\mu\nu,}_{\ \ \alpha\beta}+
   (k_2-k_1)^2\eta_{\alpha\beta}I^{\mu\nu,}_{\ \ \gamma\delta}\right)
 .\nonumber\end{eqnarray}

Putting together Eqns.~(\ref{eq:contr-a}~--~\ref{eq:tau3}),
we arrive at a simplified expression
\begin{equation}\label{eq:a-part}
{{\cal A}^{rad}_{(a)\mbox{\scriptsize
IR}}=-\kappa\,\frac{k_4^\mu\overline{\epsilon_{\mu\nu}
(k,\lambda)}k_4^\nu}{(k_4+k)^2+i\epsilon}\,{\cal
A}^{tree}(1,2,3,4)+\frac{\kappa\,{\cal O}(k)}{k\cdot
k_4+i\epsilon}}
 \end{equation}
where the ``IR'' index emphasizes that we keep only
the leading term when $k\to0$. We observe indeed that the
${\cal O}\left(\frac{1}{k}\right)$ term is proportional to the
Born amplitude without flipping any of the hard particles' spins.

Now we add on the contribution
from Fig.~\ref{fig:gggraphs}b,c,d. The result is
\begin{equation}\label{eq:irggf}
{{\cal A}^{rad}_{\mbox{\scriptsize IR}}=-\kappa\,{\cal
A}^{tree}(1,2,3,4)\,\sum_{n=1}^4{\frac{k_n^\mu
 \overline{\epsilon_{\mu\nu}(k,\lambda)}k_n^\nu}
 {(k_n+\eta_nk)^2+i\epsilon}+\frac{\kappa\,{\cal
O}(k)}{k\cdot k_n}}.}
 \end{equation}

Next we need to square this amplitude and sum over the soft
graviton spin:
\begin{equation}\label{eq:m2rd}
{\sum_\lambda{|{\cal A}^{rad}_{\mbox{\scriptsize
IR}}|^2}=\kappa^2\,|{\cal A}^{tree}|^2\,
 \sum_{i,j=1}^4\frac{k_i^\mu
k_i^\nu\,\Pi_{\mu\nu,\alpha\beta}(k)
 \,k_j^\alpha k_j^\beta}{(k_i+\eta_ik)^2\,(k_j+\eta_jk)^2},}
 \end{equation}
where the sum over graviton polarization tensors is
\begin{equation}\label{eq:polsum}{\Pi^{\mu\nu,\alpha\beta}
(k)\equiv\sum_\pm{\epsilon_\pm^\mu(k)\epsilon_\pm^\nu(k)
\overline{\epsilon_\pm^\alpha(k)}
\overline{\epsilon_\pm^\beta(k)}}=\frac{1}{2}\left(
 \Pi^{\mu\alpha}\Pi^{\nu\beta}+
\Pi^{\mu\beta}\Pi^{\nu\alpha}-
  \Pi^{\mu\nu}\Pi^{\alpha\beta}\right)}\end{equation}
and
\begin{equation}\label{eq:pi12}{\Pi^{\mu\nu}(k) =
k^\mu\lambda^\nu+k^\nu\lambda^\mu-(k\cdot\lambda)\,
\eta^{\mu\nu}}\end{equation} with an arbitrary vector
$\lambda$, {\it same} for all terms in the sum, chosen as
\mbox{$\lambda^\mu=(1,{\bf 0})$}.

Doing the algebra in the above formula gives us
\begin{equation}\label{eq:ggM}
{\sum_\lambda
{|{\cal A}^{rad}_{\mbox{\scriptsize
IR}}|^2}=\frac{\kappa^2\,|{\cal A}^{tree}|^2}{4k^2}
 \sum_{ij}{\eta_i\eta_jE_iE_j}\,\frac{(\cos{\gamma_{ij}}
 -\cos{\alpha_i}\cos{\alpha_j})^2
 -\frac{1}{2}\sin^2{\alpha_i}\sin^2{\alpha_j}}{
 (1-\cos{\alpha_i})(1-\cos{\alpha_j})},}
 \end{equation}
where $k$ now stands for the energy of the soft gluon (not the four-momentum), and
$\gamma_{ij}$ is the angle between the $(d-1)$-dimensional momenta of the hard gravitons,
$\alpha_i$~is the angle between the $i^{th}$ hard and the soft gravitons; $E_i$ is
the CM energy of the $i^{th}$ graviton and $\eta_i=+1\,(-1)$
for incoming (outgoing) hard gravitons.

At this point we make a comment on how dimensional
regularization works. In the one-loop amplitude we find a
$\frac{1}{\epsilon}$ divergence in dimensional regularization.
As pure gravity is one-loop finite and all the divergences in
one-loop graviton-graviton scattering come from the pure
gravity part only, all of this $\frac{1}{\epsilon}$ should be of
infrared origin and consequently be canceled by the square of
the amplitude ${\cal A}^{rad}$. However, ${\cal A}^{rad}$
itself is a tree level amplitude which does {\it not} diverge; the
canceling $\frac{1}{\epsilon}$ factor comes from the phase
space integral. One might wonder then why we are not getting
too much divergence: the leading term is $\frac{1}{k^2}$, so
the phase space integral introduces
\begin{equation}\label{eq:phsp1}
\int{\frac{d^{d-1}k}{k\,k^2}}
\end{equation}
which is logarithmically divergent. In the same time the
angular integration is also divergent and we find that
dimensional regularization {\it does not handle correctly} an
integral of the type
\begin{equation}\label{eq:dimbad}
{\oint\frac{d\Omega_{d-2}({\bf n})}{(1-\cos{\alpha})^2}
\sim B(1-\epsilon,-1-\epsilon)}
\end{equation}
(here $\alpha$ is the angle between the direction of ${\bf n}$
and a fixed direction.) The above Euler function is
\begin{equation}\label{eq:eul}
{B(1-\epsilon,-1-\epsilon)=
\frac{\Gamma(1-\epsilon)\Gamma(-1-\epsilon)}
{\Gamma(-2\epsilon)}\to-2+{\cal O}(\epsilon)}
\end{equation}
{\it finite}, although the integral
includes a severe collinear singularity. Fortunately, in our case,
we will not encounter this problem: the spins ``conspire'' so
that there is an additional angular factor which takes away all
collinear singularities in this integral. In other models,
however, like one with elementary massless scalars, this might
be a problem which requires further treatment.

Now we calculate the differential cross section in
$d=4-2\epsilon$ dimensions. We focus on the infrared region
only, integrating up to a cutoff $\Lambda\ll\sqrt s$
and neglecting momenta of order $\Lambda$ and above.  Such
soft graviton radiation should (and will) be sufficient to cancel
the IR divergences due to one-loop integrals. In particular, we
do not consider hard collinear gravitons.  The divergences due
to hard collinear graviton radiation (i.e. when one of the
$\alpha_i$'s is small) are not canceled by loops.  However,
these divergences are all proportional to $\Lambda$ so can be
unambiguously separated from the soft divergences. Some
rather tedious algebra leads to an integral over the direction
{\bf m} of the soft Bremsstrahlung graviton
\begin{eqnarray}\label{eq:cross-gg}
\frac{d\sigma^{rad}_{\mbox{\scriptsize
IR}}}{d\Omega_{d-1}( {\bf n})}&=&
 \frac{\kappa^2\,|{\cal A}^{tree}|^2}{(2\pi)^{3d-7}
 2^{2d+2}}\,\sum_{ij}\eta_i\eta_j\int_0^\Lambda
  \frac{dk}{k}k^{d-4}\\
  &&\times\oint{d\Omega_{d-1}({\bf m})
  \frac{(\cos{\gamma_{ij}}-\cos{\alpha_i}\cos{\alpha_j})^2
  -\frac{1}{2}\sin^2{\alpha_i}\sin^2{\alpha_j}}
  {(1-\cos{\alpha_i})(1-\cos{\alpha_j})}}.\nonumber
\end{eqnarray}
The $k$ integral has a $\frac{1}{\epsilon}$ infrared
divergence. All divergences that are collinear and infrared
simultaneously should come from the second integral.
However, we observe that the numerator in the angular integral
vanishes when the denominator does, actually canceling out the
singularity. This fact is necessary to allow us to consistently
separate collinear divergences from soft ones. In order to find
the divergent part, we need to calculate only the leading term in
\begin{eqnarray}\label{eq:angint}
{\cal F}^{(0)}(\gamma) + \epsilon{\cal F}^{(1)}
(\gamma) + \ldots\ =\hspace{40ex}&&\\
 \oint d\Omega_{d-1}({\bf m})\
  \frac{ (\cos{\gamma_{ij}}-\cos{\alpha_i}\cos{\alpha_j})^2
         -\frac{1}{2}\sin^2{\alpha_i}\sin^2{\alpha_j} }
       { (1-\cos{\alpha_i})(1-\cos{\alpha_j}) }.&&
\nonumber\end{eqnarray}
Substituting this into Eqn.~(\ref{eq:cross-gg}) we find
\begin{eqnarray}\label{eq:cross-gg1}
\frac{d\sigma^{rad}_{\mbox{\scriptsize
IR}}}{d\Omega_{d-1}({\bf n})}&=&
 -\frac{\kappa^2\,|{\cal
A}^{tree}|^2}{(2\pi)^{5-6\epsilon}2^{11-4\epsilon}}
\sum_{ij} {\cal F}^{(0)}(\gamma_{ij})\,\eta_i\eta_j\\
&&\ \times\
\left[\frac{1}{\epsilon}-2\ln{\Lambda}+6\ln{(2\pi)}+4\ln{2}+
  \frac{{\cal F}^{(1)}(\gamma_{ij})}{{\cal
F}^{(0)}(\gamma_{ij})}\right].\nonumber
\end{eqnarray}

\noindent The result in four dimensions is
\begin{equation}\label{eq:angint4}
{{\cal F}^{(0)}(\gamma)=
4\pi\left[\frac{3+\cos{\gamma}}{6}-
(1-\cos{\gamma})\ln{\frac{2}{1-\cos{\gamma}}}\right].}
\end{equation}
With this, we finally find for the cross section
\begin{eqnarray}\label{eq:cross-ggfin}
\frac{d\sigma^{rad}_{\mbox{\scriptsize
IR}}}{d\Omega}&=&
- \frac{\kappa^2\,|{\cal A}^{tree}|^2}{2^7(2\pi)^4}\,
\left(\frac{t}{s}\,\ln{\frac{-t}{s}}+\frac{u}{s}\,
\ln{\frac{-u}{s}}\right)\\
&&\times\
\left[\frac{1}{\epsilon}-2\ln{\Lambda}+6\ln{(2\pi)}+4\ln{2}+
  \frac{\sum_{ij}\eta_i\eta_j{\cal
F}^{(1)}(\gamma_{ij})}{\sum_{ij}\eta_i\eta_j{\cal
F}^{(0)}(\gamma_{ij})}\right]\,
 +{\cal O}\left(\frac{\Lambda}{\sqrt{s}}\right).\nonumber
\end{eqnarray}

\noindent We have found that the infrared divergent part is
indeed proportional to the square of the Born amplitude.
Because in the $(++;--)$ and $(++;+-)$ helicity cases there is
no IR divergence to cancel, the vanishing of the Born terms
makes sure none emerges in the radiative process.

In the $(++;++)$ helicity case we need to use the cross section
formula
\begin{equation}\label{eq:sigl0}
{\left(\frac{d\sigma[gg\to gg]}{d\Omega}\right)_{nonrad}=
 \frac{2Re(\overline{{\cal A}^{tree}}{\cal A}^{1-loop})}
 {(2\pi)^22^5s}}\end{equation}
in order to calculate the ${\cal O}(\kappa^6)$ contribution to
the cross section (see Fig.~\ref{fig:gggen}.)  Using the
Dunbar-Norridge~\cite{DunNor} 1-loop amplitude amplitude,
Eqn.~(\ref{eq:2}), we find the ${\cal O}(\kappa^6)$
contribution to the cross section for the $2\to2$ process:
\begin{eqnarray}\label{eq:crs-loop}
\left(\frac{d\sigma(++;++)}{d\Omega}\right)_{nonrad}&=&
 \ \frac{\kappa^2\,|{\cal A}^{tree}|^2}{2^7(2\pi)^4}\,
\times\\&&\times\left\{
\left(\frac{t}{s}\,\ln{\frac{-t}{s}}+\frac{u}{s}\,
\ln{\frac{-u}{s}}\right)\rule{0pt}{2em}\right.
\left(\frac{1}{\epsilon}+\ln4\pi-\ln s-\gamma
\right)\nonumber\\
&&\hspace{2em}+\left[\ln\frac{-t}{s}\ln\frac{-u}{s}+
\frac{tu}{2s^2}\,f\left(\frac{-t}{s},\frac{-u}{s}
\right)\right]
\left.\rule{0pt}{2em}\right\}.\nonumber
\end{eqnarray}

We observe that the $\frac{1}{\epsilon}$ divergence
cancels when we add together Eqns.~(\ref{eq:cross-ggfin})
and~(\ref{eq:crs-loop}). The finite term
in~Eqn.~(\ref{eq:crs-loop}) contains an undetermined scale
due to the logarithm of $s$. The occurrence of such a scale is a
common feature of dimensional regularization in the presence
of infinities. This scale is provided by the ``ultraviolet'' cutoff
in the radiative cross section. Our final result for the sum of the
cross sections~is
\begin{eqnarray}\label{sum-crs}
&&\hspace{-3em}\left(\frac{d\sigma}{d\Omega}\right)_{tree}
 + \left(\frac{d\sigma}{d\Omega}\right)_{rad.}
  +\left(\frac{d\sigma}{d\Omega}\right)_{nonrad.}=\\
&=& \frac{\kappa^4 s^5}{2048\pi^2 t^2 u^2}\,
\left\{ \rule{0pt}{2.6em}\right.
1 + {\kappa^2 s \over 16 \pi^2}
\left[\rule{0pt}{2em}\right.
 \ln\frac{-t}{s}\ln\frac{-u}{s}+
\frac{tu}{2s^2}\,f\left(\frac{-t}{s},\frac{-u}{s}\right)
\nonumber\\
&&-\left(\frac{t}{s}\,\ln{\frac{-t}{s}}+\frac{u}{s}\,
\ln{\frac{-u}{s}}\right)
\left(
3\ln(2\pi^2)+\gamma+\ln\frac{s}{\Lambda^2}+
  \frac{\sum_{ij}\eta_i\eta_j{\cal
F}^{(1)}(\gamma_{ij})}{\sum_{ij}\eta_i\eta_j{\cal
F}^{(0)}(\gamma_{ij})}
\right)
\left.\rule{0pt}{2em}\right]
 \left.\rule{0pt}{2.6em}\right\}.\nonumber
\end{eqnarray}
In this form all divergences are canceled and all logarithms are
dimensionless. There is a logarithmic dependence on the scale
where we cut off the non-infrared radiative gravitons. We
remind the reader that the finite functions
$f\left(\frac{-t}{s},\frac{-u}{s}\right)$ and $
\frac{\sum_{ij}\eta_i\eta_j{\cal
F}^{(1)}(\gamma_{ij})}{\sum_{ij}\eta_i\eta_j{\cal
F}^{(0)}(\gamma_{ij})}$ are respectively given
in~Eqn.~(\ref{eq:f}) and can be extracted from
~Eqn.~(\ref{eq:angint}).
\section{Conclusions}

This has been the first explicit investigation of the infrared
properties of a one loop amplitude in quantum gravity. We
have achieved our goal in demonstrating that the effective
theory of gravitation is not plagued by infrared divergences, its
soft divergences even cancel in the case of one-loop
graviton-graviton scattering, and also demonstrated that,
similarly to the case of QED, summation over degenerate
states in the final state suffices to get a final and sensible cross
section.

The result for graviton-graviton scattering to one-loop order is
beautiful and significant because it forms a low energy
theorem for quantum gravity. No matter what the high energy
theory of gravity may turn out to be, and independent of the
massive particles in the theory, as long as the low energy limit
leads to general relativity the scattering rate must have the
model independent form shown in~Eqn~(\ref{sum-crs}). As
expected, the quantum effective field theory of general
relativity is well-behaved in the infrared.

\end{document}